\documentclass{nature}

\usepackage{amssymb}
\usepackage{txfonts}
\usepackage{amssymb}
\usepackage{graphicx}
\usepackage{dcolumn}
\usepackage{bm}

\begin{document}

\title{On-chip coherent conversion of photonic quantum entanglement between different degrees of freedom}

\author{Lan-Tian Feng$^{1,2\dagger}$, Ming Zhang$^{3\dagger}$, Zhi-Yuan Zhou$^{1,2\dagger}$, Ming Li$^{1,2}$, Xiao Xiong$^{1,2}$, Le Yu$^{1,2}$, Bao-Sen Shi$^{1,2}$, Guo-Ping Guo$^{1,2}$, Dao-Xin Dai$^{3\star}$, Xi-Feng Ren$^{1,2\ast}$ and Guang-Can Guo$^{1,2}$ }

\maketitle
\begin{affiliations}
\item
Key Laboratory of Quantum Information, University of Science and Technology of China, CAS, Hefei, 230026, People's Republic of China

\item
Synergetic Innovation Center of Quantum Information $\&$ Quantum
Physics, University of Science and Technology of China, Hefei, Anhui
230026, China

\item
State Key Laboratory for Modern Optical Instrumentation, Centre for Optical and Electromagnetic Research,
Zhejiang Provincial Key Laboratory for Sensing Technologies, Zhejiang University, Zijingang Campus, Hangzhou
310058, China.
\\$^\dagger$These authors contributed equally to this work.
\\$^\star$e-mail: dxdai@zju.edu.cn
\\$^\ast$e-mail: renxf@ustc.edu.cn
\end{affiliations}
\begin{abstract}
In the quantum world, a single particle can have various degrees of freedom to encode the quantum information. Controlling multiple degrees of freedom simultaneously is necessary to describe a particle fully and, therefore, to use it more efficiently. Here, we introduce the transverse waveguide mode degree of freedom to quantum photonic integrated circuits, and demonstrate the coherent conversion of photonic quantum state between path, polarization and transverse waveguide mode degree of freedom on a single chip. The preservation of quantum coherence in these conversion processes is proven by single photon and two photon quantum interference using a fiber beam-splitter or on-chip beam-splitters. These results provide us with the ability to control and convert multiple degrees of freedom of photons for the quantum photonic integrated circuit-based quantum information process.
\end{abstract}

\section*{INTRODUCTION}

Compared with free space and fibre optics, photonic integrated circuits (PICs) have attracted considerable attention owing to their small footprint, scalability, reduced power consumption, and enhanced processing stability; thus, many quantum computation and quantum simulation protocols have been realized on quantum PICs (QPICs) \cite{Obrien08,Obrien09,sansoni,crespi10,spring,crespi,broome,tillmann,metcalf}. Regarding recent investigations of QPICs, polarization entanglement and path entanglement are usually used. Polarization entanglement is quite important because the polarization of photons is easy to control in free-space and various proof of principle demonstration of quantum computation schemes were carried out based on the polarization degree of freedom \cite{kok,krischek,walther}. For the path encoding method, it is possible to establish a higher-dimensional Hilbert space \cite{spring}, which means that we can encode more information per photon and increase the security of quantum systems \cite{bennett,you,bennett92,ekert}. However, the path encoding method is not easy to scale up for practical implementations.

Recently, transverse waveguide modes were introduced as a new information encoder, and they were used in multi-core \cite{randel} and few-mode fibres \cite{richard}, as well as in classical PICs \cite{dai12}, to satisfy the increasing demand for the capacity of optical interconnects. A multi-mode waveguide can support many transverse waveguide modes, which form a set of orthogonal basis for the transverse spatial distribution of energy; thus, it is appropriate for carrying more information using photons. For example, the eight channel information encoding process has been demonstrated in a $2.363 \mu m$ wide multi-mode waveguide \cite{dai14}. This degree of freedom may have great potential in quantum optics, such as realizing high-dimensional quantum operation, maintaining the polarization entanglement resource in a high-birefringence integrated device.

We can use multiple degrees of freedom of a quantum particle simultaneously, which will certainly increase the information capacity of qubits \cite{Kwait1997,barre,walborn,gauth}.  As for photons, polarization, frequency, time, orbital angular momentum, and even the transverse mode entanglement have been used in free-space quantum systems, and some of them have been used in fibre quantum systems \cite{jeremy09}. For example, quantum teleportation of the composite quantum states of a single photon encoded in both spin and orbital angular momentum was recently demonstrated in free-space \cite{pan}. Experimental realization of multi-degree of freedom entanglement poses significant challenges to the coherent control of multiple degrees of freedom simultaneously and to realizing quantum logic gates between independent qubits of different degrees of freedom.

In the following, to show the potential utility of transverse waveguide modes in the quantum information process, we demonstrate that quantum coherence is preserved when photons in different transverse waveguide modes propagate in a multi-mode waveguide. We will also show the coherent on-chip conversion of quantum states between different degrees of freedom, such as path, polarization, and the transverse waveguide mode. Here, coherent conversion refers to the preservation of coherence of quantum state, including the indistinguishability between the single photons, the stability of relative phase of superposition state and entangled state in the processes of photon transmission and conversion between different degrees of freedom.

\section*{RESULTS}
\subsection{Experimental setup}

The transverse waveguide modes discussed in this study are the three lowest-order modes, i.e., $\mathrm{TE}_0$, $\mathrm{TE}_1$ and $\mathrm{TM}_0$ in a multi-mode waveguide, as shown in upright inset of Figure 1a. Here, we use a silicon-on-insulator (SOI) strip waveguide with a cross section of $\sim750 nm\times220 nm$. To effectively and accurately excite and manipulate these transverse waveguide modes in this multi-mode waveguide, the low-loss and low-crosstalk mode (de)multiplexer is one of the most important devices. In this work, we choose two different structures for the conversion between different degrees of freedom. The first one is a special polarization-dependent mode converter, which can convert laser beam in $\mathrm{TE}$ and $\mathrm{TM}$ polarizations into the $\mathrm{TE}_0$ and $\mathrm{TE}_1$ modes \cite{dai11,dai112}, respectively, as shown in Supplementary Fig. 1a. The other one is a mode multiplexer, which can convert laser beam in path 1 and path 2 into the $\mathrm{TE}_0$ and $\mathrm{TE}_1$ waveguide modes \cite{dai13}, respectively, as shown in Supplementary Fig. 1b. Although these elements work well for laser beam, they have not yet been used for quantum signals. After conversion, the $\mathrm{TE}_0$ and $\mathrm{TE}_1$ mode photons propagate in the multi-mode waveguide for a certain distance and then convert back to photons with different polarizations or different optical paths, as shown in Figure 1b.

To test the coherent property of the two photons after undergoing different conversion processes, a Hong-Ou-Mandel (HOM) interferometer is used. HOM interference is a basic type of quantum interference that reflects the bosonic properties of a single particle and is generally used to test the quantum properties of single qubits \cite{Hong}. It can be described as follows: when two indistinguishable photons enter a 50/50 beam splitter (BS) from different sides at the same time, the two photons will come out together and never be in different output ports. Experiments typically control the arrival time of two photons by adjusting the path-length difference between them and measure the photon coincidence (the case in which two photons arrive at two detectors simultaneously) of the two output ports of the BS. When two indistinguishable photons completely overlap at the BS, they give rise to the maximum interference effect and no coincidence exists. Visibility is defined as $V_{1}=(C_{\mathrm{max}}-C_{\mathrm{min}})/C_{\mathrm{max}}$, where $C_{\mathrm{max}}$ is the maximum coincidence and $C_{\mathrm{min}}$ is the minimum coincidence. For perfect quantum interference, $C_{\mathrm{min}}=0$ and $V_{1}=1$. Or we can collect the photons from one output port of the BS, send them to the second 50/50 BS, and then measure the coincidence \cite{Rarity}. In this case, a peak will be observed and the visibility is modified as follows: $V_{2}=(C_{\mathrm{max}}-C_{\mathrm{min}})/C_{\mathrm{min}}$. For perfect quantum interference, $C_{\mathrm{max}}=2C_{\mathrm{min}}$ and $V_{2}=1$.

The experimental setup is shown in Figure 2. The degenerate $1558 nm$ photon pair source is generated using a type-II phase-matched periodically poled potassium titanyl phosphate ($\mathrm{KTiOPO}_{4}$) crystal in a Sagnac interferometer pumped by a continuous-wave $779 nm$ laser\cite{zhou}, which operates in a single circulation direction. Each photon is coupled into a single mode fibre and then sent to a fibre array. In one arm, the fibre coupler is mounted on a one dimensional translator with a step of $10\mu m$. By moving the translator, we can modify the arrival time of the single photons on BS and thus observe the HOM interference effect. The grating coupling method is used to couple the single photons into/out of the chip from/into the fibre arrays. The output photon pairs collected by the second fibre array are sent into a fibre BS to produce the HOM interference.

The indistinguishability of the photon pairs from the source is characterized using a standard HOM interferometer with a fibre BS. The dip represents the quantum interference of two photons, and the coherence length of the photons determines its width, as shown in Supplementary Fig. 2. Here, we obtain a raw visibility of $96.3\pm2.8\%$ ($96.8\pm2.8\%$ with background subtraction) and an optical coherence length of $448.7\pm19.8 \mu m$. The deviation of the visibility from $100\%$ is attributed to the polarization distortion of the photons during the propagation in the fibre, the photon source variability, or both.

\subsection{Single photon state conversion}

The coherent conversion of single photon state between different degrees of freedom are tested at first. The sketch map and CCD picture of the first sample are shown in Figure 3a and 3b, respectively (see Supplementary Fig. 3 for detail). Two single photons with orthogonal polarizations from the fibre array are coupled into different single mode waveguides, i.e., the $\mathrm{TE}_0$ and $\mathrm{TM}_0$ mode, by a $\mathrm{TE}$-type grating and a $\mathrm{TM}$-type grating, respectively. Then, the two single mode waveguides combine together with a polarization beam-splitter (PBS) based on a bent directional coupler \cite{dai11}. A special mode converter based on an adiabatic taper is cascaded and polarization-dependent mode conversion happens, as reported previously \cite{dai112}. The $\mathrm{TM}_0$ mode is then converted into the $\mathrm{TE}_1$ mode after propagating along this adiabatic taper while there is no mode conversion for the $\mathrm{TE}_0$ mode. As a result, photons with different polarizations are converted to different transverse waveguide modes (i.e., the $\mathrm{TE}_0$ and $\mathrm{TE}_1$ modes). This is experimentally proven by a near-field scanning optical microscope (NSOM), which can measure the evanescent field distribution of the guided-mode in an optical waveguide. The results are shown in Supplementary Fig. 4, which clearly prove the function of the PBS and the mode converter. The two photons in the $\mathrm{TE}_0$ mode and the $\mathrm{TE}_1$ mode propagate along the multi-mode waveguide for $870 \mu m$ and are converted back to different polarizations by a similar mode converter structure. We collect the two output photons from the two different gratings and then send them to the HOM test system. By moving the one dimensional translator, we can control the arrival time difference between the two photons and thus get the relation between the coincidence and the path length difference. Figure 3e gives the measured HOM interference between the two photons undergoing different conversion processes. The raw visibility is $92.3\pm 5.0\%$ ($94.8\pm5.0\%$ with background subtraction), and the coherent length is $458.7\pm37.8 \mu m$, which proves unambiguously the preservation of quantum coherence during the conversion from the polarization to the transverse waveguide mode and back to polarization process.

Then, we tested the single photon state conversion between the optical path and transverse waveguide mode degrees of freedom with sample 2, as shown in Figure 3c and 3d (see Supplementary Fig. 5 for detail). In this case, two photons with the same polarization from the fibre array are coupled into different single mode waveguides, both having the $\mathrm{TE}_0$ mode, by two $\mathrm{TE}$-type gratings, respectively. Then, the two single mode waveguides combine together with a mode multiplexer such that one $\mathrm{TE}_0$ mode is converted to the $\mathrm{TE}_1$ mode in the bus waveguide while the other $\mathrm{TE}_0$ mode is kept unchanged. Thus, photons in different optical paths are converted to different transverse waveguide modes. The two photons in the $\mathrm{TE}_0$ mode and the $\mathrm{TE}_1$ mode propagate along the multi-mode waveguide for $30 \mu m$, and then are converted back to the $\mathrm{TE}_0$ mode in two distinct single mode waveguides with a mode demultiplexer. Unlike the first sample, HOM interference occurs on an on-chip BS (3 dB coupler). Figure 3f gives the result with a raw visibility of $96.0\pm3.3\%$ ($97.3\pm3.3\%$ with background subtraction), and the coherent length is $460.2\pm28.1 \mu m$, which illustrates the coherent conversion of quantum signals from path encoding to transverse waveguide mode encoding.

\subsection{Quantum entanglement conversion}

Coherent conversion of the quantum entangled state is also proven in our experiment using the third and fourth samples shown in Figure 4a-4d (see Supplementary Figs. 6 and 7 for detail), respectively. For the third sample, two single photons with the same polarization are coupled to single mode waveguides by two gratings, respectively. Then, they interfere at the BS and generate a two-photon quantum NOON state \cite{giovan}, encoded on the path as $(\left|2\right\rangle_0\left|0\right\rangle_1 -\left|0\right\rangle_0\left|2\right\rangle_1)/\sqrt{2}$, where $\left|n\right\rangle_i$ denotes $n$ photons in path $i$, for $n=0,1,2$, and $i=0,1$. With a mode multiplexer, the state will be changed to $(\left|2\right\rangle_{\mathrm{TE}_1}\left|0\right\rangle_{\mathrm{TE}_0} -\left|0\right\rangle_{\mathrm{TE}_1}\left|2\right\rangle_{\mathrm{TE}_0})/\sqrt{2}$ because the photons in path 0 will be in the $\mathrm{TE}_1$ mode while photons in path 1 will be in the $\mathrm{TE}_0$ mode. After propagation in the multi-mode waveguide for a distance of $30 \mu m$, this transverse waveguide mode two-photon NOON state is changed back to a path NOON state. Two-photon interference, or two-photon NOON state interference, is measured by using the second on-chip BS, as shown in Figure 4e (red dots). The phase between the two arms is adjusted by using a thermal-tuning method. Classical interference is also measured for comparison with a visibility of $99.9\pm7.8\%$ (black dots). We observe a entangled state interference visibility of $90.3\pm7.8\%$ ($94.0\pm8.2\%$ with background subtraction) with a period (heater power $32.5\pm0.7 mW$) approximately half of the classical interference (heater power period is $66.8\pm2.8 mW$).

Finally, to show that these conversion processes can be cascaded, we combine several structures together on a single chip, and a two-photon quantum NOON state is converted between these three degrees of freedom. The sample is shown in Figures 4c and 4d. Two single photons with the same polarization are coupled to single mode waveguides and then interfere at the on-chip BS, generating a quantum entangled state encoded on the path as $(\left|2\right\rangle_0\left|0\right\rangle_1 -\left|0\right\rangle_0\left|2\right\rangle_1)/\sqrt{2}$. It is firstly converted to a transverse waveguide mode NOON state $(\left|2\right\rangle_{\mathrm{TE}_1}\left|0\right\rangle_{\mathrm{TE}_0} -\left|0\right\rangle_{\mathrm{TE}_1}\left|2\right\rangle_{\mathrm{TE}_0})/\sqrt{2}$ and then changed to $(\left|2\right\rangle_{\mathrm{TM}_0}\left|0\right\rangle_{\mathrm{TE}_0}-\left|0\right\rangle_{\mathrm{TM}_0}\left|2\right\rangle_{\mathrm{TE}_0})/\sqrt{2}$, which is a polarization NOON state. To show that the whole process is actually the same as what we described above, it is important to show that the two photons at the output are either both in $H$ polarization or in $V$ polarization and never have different polarizations. We collect the output photons from each grating coupler and conduct the HOM interference test with a fibre BS. As shown in Figures 4f and 4g, peaks are observed as the two photons arrive at the on-chip BS simultaneously. The raw visibilities are $96.8\pm7.8\%$ ($98.2\pm7.9\%$ with background subtraction) and $96.7\pm8.3\%$ ($98.3\pm8.5\%$ with background subtraction), and the coherent lengths are $446.1\pm37.0\mu m$ and $407.7\pm31.2\mu m$, respectively. This means that when the two-photon path NOON state is generated, there are two photons at output 1 or at output 2, as we predicted.

\section*{DISCUSSION}

We conclude that our experiment demonstrates unambiguously the coherent propagation of quantum signals encoded on transverse waveguide modes and the on-chip coherent conversion of quantum entanglement between different degrees of freedom. Although only two lower transverse waveguide modes are discussed, this newly introduced degree of freedom shows us the possibility of encoding quantum information within a higher-dimensional Hilbert space, which is useful for the investigation of the on-chip high dimensional quantum information process, such as teleportation using qudits \cite{bennett,you}, quantum dense coding \cite{bennett92}, and quantum key distribution \cite{ekert}. For example, a two-photon three dimensional path entangled state can be converted to a three dimensional transverse waveguide mode entangled state easily using our mode converters (similar demonstration between path and orbital angular momentum was recently realized in free space \cite{Fickler}).

The on-chip coherent conversion of quantum entangled state that are encoded onto path, polarization and the transverse waveguide mode, shows us the ability to control these degrees of freedom, which has great potential in on-chip hyper-entangled quantum systems. Hyper-entanglement \cite{Kwait1997,Cinelli}, where qubits are entangled in two or more degrees of freedom, has shown advantages in quantum information applications. Using hyper-entanglement will make it much easier to perform quantum logic gates \cite{Fio}, and will also enable new capabilities in quantum information process, such as remote preparation of entangled states, full Bell-state analysis, and improved super-dense coding \cite{barre,walborn}, as well as the possibility of quantum communication with larger alphabets \cite{gauth}. Path-polarization hyper-entangled and cluster states of photons on a chip was recently realized \cite{Ciampini}.

The chips we used are based on SOI waveguides, which have been developed well and used widely because of the CMOS compatibility and the ultra-small footprint. The operation wavelength of the silicon chip lies in the whole telecom band, which is compatible with present fibre communication networks. Quantum information processing based on silicon photonic chip will be a good candidate of quantum processor for quantum communication networks. Note that because different transverse waveguide modes have different effective refractive indices, relative phases will be generated when photons in different transverse waveguide modes propagate along the multi-mode waveguide. For the on-chip quantum information process, this phase can be adjusted by using a thermal-tuning method. In our experiment, all the measurements of the HOM interference were performed on the path degree of freedom for the sake of simplification.

While this report was being written, quantum interference between transverse waveguide modes was realized \cite{Mohan}.

\section*{METHODS}

\noindent {\bf Sample design and fabrication}

The chip includes some key components, including the PBSs, the mode multiplexers and the mode converters. All the components are designed according to the optical waveguide theory and the coupled-mode theory. The simulation tools include the Fimmprop (PhotoDesign, Oxford, UK) employing an eigenmode expansion and the matching method and Lumerical software (Lumerical Solutions, Inc. London, UK) with the three-dimensional time-domain finite-difference (3D-FDTD) method.
The PBS is designed with a bent directional coupler consisting of two parallel bent waveguides \cite{dai11}. These two bent waveguides have different core widths and could be designed to satisfy the phase-matching condition for the coupling of TM polarization; consequently, TM polarized light could be coupled to the cross port completely when choosing the length of the coupling region appropriately. On the other hand, for $\mathrm{TE}$ polarization, the phase-matching condition is not satisfied because of the birefringence of the waveguides. Thus, $\mathrm{TE}$ polarized light goes through without any significant coupling. In this way, $\mathrm{TE}$- and $\mathrm{TM}$- polarized light are separated within a very short length that is close to the coupling length of TM polarization.
The mode multiplexer is designed with an asymmetric directional coupler, which consists of a narrow access waveguide close to the wide bus waveguide\cite{dai13}. The widths of the narrow access waveguide and the wide bus waveguide are chosen optimally to satisfy the phase matching condition so that the fundamental mode of the narrow access waveguide can be coupled to the first higher-order mode in the bus waveguide completely. On the other hand, there is almost no coupling from the fundamental mode of the wide bus waveguide to any modes in the narrow access waveguide.
The mode converter is designed with an adiabatic taper based on SOI strip waveguides with an air upper-cladding \cite{dai112}. For an SOI strip waveguide whose cross section has vertical asymmetry, mode hybridization happens when choosing some special core width $w_{\mathrm{co}0}$. The hybridized modes have comparable x- and y-components for the electrical fields. Such mode hybridization will introduce a mode conversion when light propagates along an adiabatic taper structure whose end widths $w_1$ and $w_2$ are chosen such that $w_1<w_{\mathrm{co}0}<w_2$.

For the fabrication of the present sample, the process started from an SOI wafer with a $220 nm$-thick top silicon layer. An E-beam lithography process with the MA-N2403 photoresist was carried out to make the pattern of waveguides, which was then transferred to the top silicon layer via an inductively-coupled-plasma etching process. Grating couplers were made using a second etching process to achieve an efficient fibre-chip coupling.

\noindent {\bf Grating coupling method}

Grating couplers are very popular for realizing efficient coupling between the chip and fibers at the input/output ends \cite{Laere}. Here we use two types of grating couplers, i. e., $\mathrm{TE}$-type and $\mathrm{TM}$-type, which are designed for $\mathrm{TE}$- and $\mathrm{TM}$-polarized lights, respectively. In our design, the grating periods are chosen as $640 nm$ and $1040 nm$ for the $\mathrm{TE}$- and $\mathrm{TM}$-type grating couplers, respectively. To avoid reflection at the waveguide-grating interface, light is coupled in (out) at a small angle ($15^\circ$ in our experiment) with respect to the vertical direction. The peak coupling efficiencies are about $30\%$.

\noindent {\bf Photon source}

The continuous-wave pump laser at $779 nm$ is from a Ti: sapphire laser (Coherent MBR 110). It is collected into single mode fibre before entering the Sagnac-loop. A quarter wave plate (QWP) and a half wave plate (HWP) are used to control the phase and intensity of the pump beams in the Sagnac-loop. In the present experiment, the pump laser with vertical polarization is focused by a lens with a focus length of $200 mm$, whose beam waist is approximately $40 \mu m$ at the centre of the periodically poled potassium titanyl phosphate (PPKTP) crystal. The type-II PPKTP (Raicol crystals) crystal has a size of $1 mm\times2 mm\times10 mm$, with a periodical poling period of $46.2 \mu m$. The temperature of the PPKTP crystal is controlled by a homemade temperature controller with a stability of $2 mK$. After a double polarization beam splitter (DPBS), the polarization of the pump beam is changed to horizontal by a double half wave plate (DHWP) before the PPKTP crystal. The orthogonal polarized photon pairs generated in the counterclockwise direction are separated by the DPBS and collected into single mode fibres by using a lens set consisting of two lenses with different focus lengths of $100 mm$ and $50 mm$ at each output port of the DPBS, respectively. The pump beam is removed using a long pass filter (FELH1400). We use HWPs and QWPs to control the polarizations of the photon pair before injecting into the silicon chip. The output photons from the chip are detected by two InGaAs single photon avalanche detectors (D220, free running single photon detector.).

\noindent {\bf Data availability}

The authors declare that the data supporting the findings of this study are available within the article and its supplementary information files.

\section*{Acknowledgments}

This work was funded by NBRP (grant nos. 2011CBA00200 and 2011CB921200),
the Innovation Funds from the Chinese Academy of Sciences (grant no.
60921091), NNSFC (grant nos.11374289, 61590932, 61422510, 11374263, 61431166001, 61525504), the Fundamental Research Funds for the Central Universities, the Open Fund of the State Key Laboratory on Integrated Optoelectronics and the Doctoral Fund of the Ministry of Education of China (No. 20120101110094).

\section*{Author contributions}

All authors contributed extensively to the work presented in this
paper. M.Z., D.X.D prepared the samples.
L.T.F., Z.Y.Z., B.S.S. and X. F. R performed the measurements and data analysis.
M.L., X.X., L.Y., G.P.G. and G.C.G. conducted theoretical analysis.
X.F.R, D.X.D wrote the manuscript and supervised
the project.

\section*{Additional information}

\textbf{Supplementary information} is available in the online version of the paper. Correspondence and requests for materials should be addressed to X.F.R or D.X.D.

\textbf{Competing financial interests:} The authors declare no competing financial interests.

\clearpage

\newpage
\bigskip
\noindent \textbf{Figure 1} \textbf{On-chip coherent conversion of photonic quantum entanglement between different degrees of freedom.} (a) Transverse waveguide mode as a new on-chip quantum information encoder. In free space high dimensional quantum information processes, orbital angular momentums of photons are usually used to encode information. Correspondingly, transverse waveguide mode can be used as a new degree of freedom for on-chip high dimensional quantum information process. Inset on upright corner shows the energy distributions of the fundamental mode ($\mathrm{TE}_0$, $\mathrm{TM}_0$) and the first higher-order mode ($\mathrm{TE}_1$) in a multi-mode waveguide. The silicon-on-insulator (SOI) strip waveguide has a cross section of $\sim750 nm\times220 nm$. (b) On-chip coherent conversion of quantum states between different degrees of freedom, such as path, polarization, and the transverse waveguide mode, is essential for using different degrees of freedom simultaneously.

\bigskip

\noindent \textbf{Figure 2} \textbf{Experimental setup for the two-photon source and sample measurement.} The continuous-wave pump laser at $779 nm$ is from a Ti: sapphire laser (Coherent MBR 110). It is collected into single mode fibre (SMF) before entering the Sagnac-loop. A quarter wave plate (QWP) and a half wave plate (HWP) are used to control the phase and intensity of the pump beams in the Sagnac-loop. In the present experiment, the pump laser with vertical polarization is focused by a lens (L) with a focus length of $200 mm$, whose beam waist is approximately $40 \mu m$ at the centre of the periodically poled potassium titanyl phosphate (PPKTP) crystal. The type-II PPKTP (Raicol crystals) crystal has a size of $1 mm\times2 mm\times10 mm$, with a periodical poling period of $46.2 \mu m$. The temperature of the PPKTP crystal is controlled by a homemade temperature controller with a stability of $2 mK$. After a double polarization beam splitter (DPBS), the polarization of the pump beam is changed to horizontal by a double half wave plate (DHWP) before the PPKTP crystal. The orthogonal polarized photon pairs generated in the counterclockwise direction are separated by the DPBS and collected into single mode fibres by using a lens set consisting of two lenses with different focus lengths of $100 mm$ and $50 mm$ at each output port of the DPBS, respectively. The pump beam is removed using a long pass filter (LPF, FELH1400). We use HWPs and QWPs to control the polarizations of the photon pair before injecting into the silicon chip. The output photons from the chip are detected by two InGaAs single photon avalanche detectors (SPAD, D220, free running single photon detector), with polarization controlled by fibre polarization controllers(PCs). The grating coupling method is used to couple the single photons into/out of the chip from/into the fibre arrays.
\bigskip

\noindent \textbf{Figure 3} \textbf{Single photon state conversion.} (a) and (b) are the sketch map and CCD picture (scale bar $ 250 \mu m$) of the first sample, respectively. Two single photons with orthogonal polarizations from the fibre array are coupled into different single mode waveguide as $\mathrm{TE}_0$ and $\mathrm{TM}_0$ modes, respectively by a $\mathrm{TE}$-type grating and a $\mathrm{TM}$-type grating and then converted to different transverse waveguide modes, i.e., the $\mathrm{TE}_0$ and $\mathrm{TE}_1$ modes, by a mode converter after a polarization beam-splitter (PBS). After propagating along the multi-mode waveguide for $870 \mu m$, the two photons are converted back with different polarizations and are coupled out to a fibre beam-splitter (BS) to produce a Hong-Ou-Mandel (HOM) interference. Coincidence measurement is performed when adjusting the position of the one-dimensional translator. (c) and (d) are the sketch map and CCD picture (scale bar $ 250 \mu m$) of the second sample, respectively. Two single photons with the same polarization from the fibre array are coupled into different single mode waveguides, both having the $\mathrm{TE}_0$ mode, by two $\mathrm{TE}$-type gratings, respectively. Then, with a mode multiplexer, photons in different optical paths are converted to different transverse waveguide modes. After propagating along the multi-mode waveguide for 30 $\mu m$, the two photons are divided into two distinct single mode waveguides by a reversed process. HOM interference occurs on an on-chip BS (3dB coupler), which was performed on the path degree of freedom. (e) HOM interference between the two photons which undergo different conversion processes via a fibre BS for the first sample. The raw visibility is $92.3\pm 5.0\%$ ($94.8\pm5.0\%$ with background subtraction), which proves unambiguously the preservation of quantum coherence in the process of polarization to transverse waveguide mode and back to polarization. (f) HOM interference between two photons undergoing different conversion processes using an on-chip BS for the second sample. The visibility is $96.0\pm3.3\%$ ($97.3\pm3.3\%$ with background subtraction), which proves unambiguously the preservation of quantum coherence in the process of path to transverse waveguide mode and back to path. Error bar comes from the Poisson statistical distribution. The experimental data are fitted with a triangle function.

\bigskip

\noindent \textbf{Figure 4} \textbf{Quantum entangled state conversion.} (a) and (b) are the sketch map and CCD picture (scale bar $ 250 \mu m$) of the third sample, respectively. Two single photons with the same polarization are coupled to single mode waveguides by two gratings, respectively. Then, they interference at the first beam-splitter (BS, 3dB coupler) and generate a two-photon quantum NOON state, encoded on the path as $(\left|2\right\rangle_0\left|0\right\rangle_0 -\left|0\right\rangle_0\left|2\right\rangle_1)/\sqrt{2}$. With a mode multiplexer, the state will be changed to $(\left|2\right\rangle_{\mathrm{TE}_1}\left|0\right\rangle_{\mathrm{TE}_0} -\left|0\right\rangle_{\mathrm{TE}_1}\left|2\right\rangle_{\mathrm{TE}_0})/\sqrt{2}$. After propagation in the multi-mode waveguide for a distance of $30 \mu m$, this transverse waveguide mode NOON state is changed back to a path NOON state. Two-photon interference, or two-photon NOON state interference, is measured by using the second on-chip BS, as shown in (e) (red dots). Classical interference is also measured for comparison (black dots). We observe a raw interference visibility of $90.3\pm7.8\%$ ($94.0\pm8.2\%$ with background subtraction) with a period approximately half of the classical interference. (c) and (d) are the sketch map and CCD picture (scale bar $ 250 \mu m$) of the fourth sample, respectively. A quantum two-photon NOON state is generated by an on-chip BS and then converted between the three degrees of freedom (path, transverse waveguide mode and polarization). (f) and (g) are Hong-Ou-Mandel (HOM) interference patterns between the two photons from TE and TM outputs of the sample, respectively. Error bar comes from the Poisson statistical distribution. The raw visibilities are $96.8\pm7.8\%$ ($98.2\pm7.9\%$ with background subtraction) and $96.7\pm8.3\%$ ($98.3\pm8.5\%$ with background subtraction), respectively. These results prove unambiguously the preservation of quantum coherence in the conversion of quantum entangled state between different degrees of freedom. It should be noted that, all the measurements of the HOM interference were performed on the path degree of freedom for the sake of simplification.


\section*{REFERENCES}
\begin{thebibliography}{99}
\bibitem{Obrien08} Politi, A., Cryan, M. J., Rarity, J. G., Yu, S., and O'Brien, J. L. Silica-on-silicon waveguide quantum circuits. \emph{Science} \textbf{320}, 646-649 (2008).
\bibitem{Obrien09} Politi, A., Matthews J. C. F., Thompson M. G. and O'Brien J. L. Integrated quantum photonics. \emph{IEEE J. Sel. Top. Quantum Electron.} \textbf{15}, 1673-1684 (2009).
\bibitem{sansoni} Sansoni, L. \emph{et al.} Polarization entangled state measurement on a chip. \emph{Phys. Rev. Lett.} \textbf{105}, 200503 (2010).
\bibitem{crespi10} Crespi, A. \emph{et al.} Integrated photonic quantum gates for polarization qubits. \emph{Nature Commun.} \textbf{2}, 566 (2011).
\bibitem{spring}Spring, J. B. \emph{et al.} Boson sampling on a photonic chip.\emph{ Science} \textbf{339}, 798-801 (2013).
\bibitem{crespi} Crespi, A. \emph{et al.} Integrated multimode interferometers with arbitrary designs for photonic boson sampling. \emph{Nature Photon.} \textbf{7}, 545-549 (2013).
\bibitem{broome} Broome, M. A. \emph{et al.} Photonic boson sampling in a tunable circuit. \emph{Science} \textbf{339}, 794-798 (2013).
\bibitem{tillmann} Tillmann, M. \emph{et al.} Experimental boson sampling.\emph{ Nature Photon.} \textbf{7}, 540-544 (2013).
\bibitem{metcalf} Metcalf, B. J. \emph{et al.} Quantum teleportation on a photonic chip. \emph{Nature Photon.} \textbf{8}, 770-774 (2014).
\bibitem{kok} Kok, P. \emph{et al.} Linear optical quantum computing with photonic qubits. \emph{Rev. Mod. Phys.} \textbf{79}, 135-174 (2007).
\bibitem{krischek} Krischek, R. \emph{et al.} Ultraviolet enhancement cavity for ultrafast nonlinear optics and high-rate multiphoton entanglement experiments. \emph{Nature Photon.} \textbf{4}, 170-173 (2010).
\bibitem{walther} Walther, P.  Experimental one-way quantum computing. \emph{Nature} \textbf{434}, 169-176 (2005).
\bibitem{bennett} Bennett, C. H. \emph{et al.} Teleporting an unknown quantum state via dual classical and Einstein-Podolsky-Rosen channels.\emph{ Phys. Rev. Lett.} \textbf{70}, 1895-1899 (1993).
\bibitem{you} Zhan, y., Zhang, Q., Wang, Y. and Ma, P. Schemes for teleportation of an unknown single-qubit quantum state by using an arbitrary high-dimensional entangled state. \emph{Chin. Phys. Lett.} \textbf{27}, 10307-10310 (2010).
\bibitem{bennett92} Bennett, C. H. and Wiesner, S. J. Communication via one- and two-particle operators on Einstein-Podolsky-Rosen states. \emph{Phys. Rev. Lett.} \textbf{69}, 2881-2884 (1992).
\bibitem{ekert} Ekert, A. K. Quantum cryptography based on Bell's theorem. \emph{Phys. Rev. Lett.} \textbf{67}, 661-663 (1991).
\bibitem{randel} Randel, S. Integrated photonic quantum gates for polarization qubits. 6¡Á56-Gb/s mode-division multiplexed transmission over 33-km few-mode fiber enabled by 6¡Á6 MIMO equalization. \emph{Opt. Express} \textbf{19}, 16697-16707 (2011).
\bibitem{richard} Richardson, D. J., Fini, J. M., and Nelson, L. E. Space-division multiplexing in optical fibres. \emph{Nature Photon.} \textbf{7}, 354-362 (2013).
\bibitem{dai12} Dai, D. Silicon mode-(de)multiplexer for a hybrid multiplexing system to achieve ultrahigh capacity photonic networks-on-chip with a single-wavelength-carrier light. in Asia Communications and Photonics Conference, OSA Technical Digest (online) (Optical Society of America, 2012), ATh3B.3.
\bibitem{dai14} Wang, J., He, S., and Dai, D. On-chip silicon 8-channel hybrid (de)multiplexer enabling simultaneous mode- and polarization-division-multiplexing, \emph{Laser Photonics Rev.} \textbf{8}, L18-L22 (2014).
\bibitem{Kwait1997} Kwiat, P. G. Hyper-entangled states. \emph{J. Mod. Opt.} \textbf{44}, 2173-2184 (1997).
\bibitem{barre}Barreiro, J. T., Wei, T.-C. and Kwiat, P. G. Beating the channel capacity limit for linear photonic superdense coding. \emph{Nature Phys.} \textbf{4}, 282-286 (2008).
\bibitem{walborn}Walborn, S. P. Hyperentanglement: Breaking the communication barrier. \emph{Nature Phys.} \textbf{4}, 268-269 (2008).
\bibitem{gauth}Gauthier, D. J. Integrated photonic quantum gates for polarization qubits. Quantum key distribution using hyperentangled time-bin states. In Conference on Coherence and Quantum Optics (Optical Society of America, 2013), pp. W2A-2.
\bibitem{jeremy09} O'Brien, J. L., Furusawa, A. and Vu\v{c}kovi\'{c}, J. Photonic quantum technologies. \emph{Nature Photon.} \textbf{3}, 687-695 (2009).
\bibitem{pan} Wang, X. \emph{et al.} Quantum teleportation of multiple degrees of freedom of a single photon. \emph{Nature} \textbf{518}, 516-519 (2015).
\bibitem{dai13} Dai, D., Wang, J, and Shi, Y. Silicon mode (de)multiplexer enabling high capacity photonic networks-on-chip with a single-wavelength-carrier light. \emph{Opt. Lett.} \textbf{38}, 1422-1424 (2013).
\bibitem{dai11} Dai, D. and Bowers, J. E. Novel ultra-short and ultra-broadband polarization beam splitter based on a bent directional coupler. \emph{Opt. Express} \textbf{19}, 18614-18620 (2011).
\bibitem{dai112} Dai, D. and Bowers, J. E. Novel concept for ultracompact polarization splitter-rotator based on silicon nanowires. \emph{Opt. Express} \textbf{19}, 10940-10949 (2011).
\bibitem{Hong} Hong, C. K., Ou, Z. Y., and Mandel, L. Measurement of subpicosecond time intervals between two photons by interference. \emph{Phys. Rev. Lett.} \textbf{59}, 2044-2046 (1987).
\bibitem{Rarity} Rarity, J. G., Tapster, P. R. Fourth-order interference in parametric downconversion. \emph{J. Opt. Soc. Am. B: Opt. Phys.} \textbf{6}, 1221-1226 (1989).
\bibitem{zhou} Li, Y., Zhou, Z., Ding, D. and Shi, B. CW-pumped telecom band polarization entangled photon pair generation in a Sagnac interferometer. \emph{Opt. Express} \textbf{23}, 28792-28800 (2015).
\bibitem{giovan} Giovannetti, V., Lloyd, S. and Maccone, L. Quantum-enhanced measurements: beating the standard quantum limit. \emph{Science} \textbf{306}, 1330-1336 (2004).
\bibitem{Fickler} Fickler, R. \emph{et al.} Interface between path and orbital angular momentum entanglement for high-dimensional photonic quantum information. \emph{Nature Commun.} \textbf{5}, 4502 (2014).
\bibitem{Cinelli} Cinelli, C., Barbieri, M., Perris, R., Mataloni, P., and De Martini F. All-Versus-Nothing Nonlocality Test of Quantum Mechanics by Two-Photon Hyperentanglement. \emph{Phys. Rev. Lett.} \textbf{95}, 240405 (2005).
\bibitem{Fio} Fiorentino, M. and Wong, F. N. C., Deterministic controlled-not gate for single-photon two-qubit quantum logic. \emph{Phys. Rev. Lett.} \textbf{93}, 070502 (2004).
\bibitem{Ciampini} Ciampini, M. A. \emph{et al.} Path-polarization hyperentangled and cluster states of photons on a chip. \emph{Light: Science and Applications} \textbf{5}, e16064 (2016).
\bibitem{Mohan} Mohanty, A. \emph{et al.} Quantum interference between transverse spatial waveguide modes. Preprint at http://arxiv.org/abs/1601.00121v1(2016).
\bibitem{Laere} Van Laere, F. \emph{et al.} Compact focusing grating couplers for silicon-on-insulator integrated circuits. \emph{IEEE Photonic Tech L} \textbf{19}, 1919-1921, (2007).

\end{thebibliography}
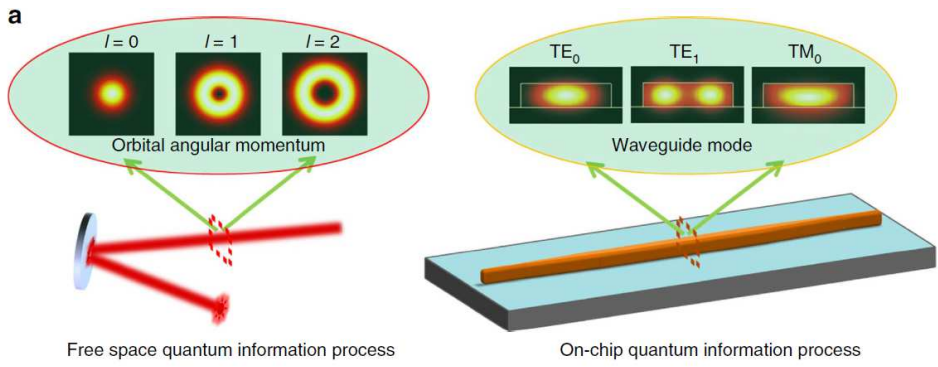
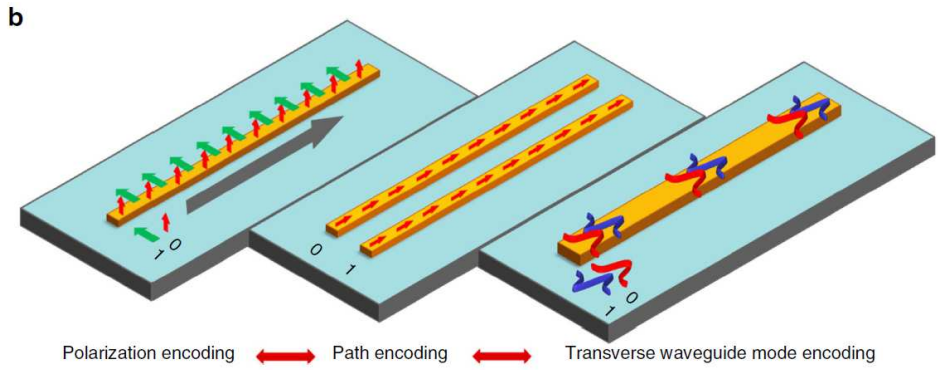
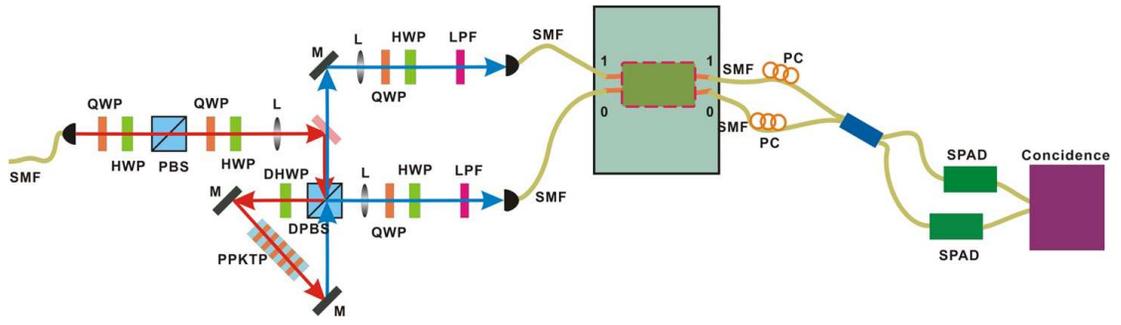
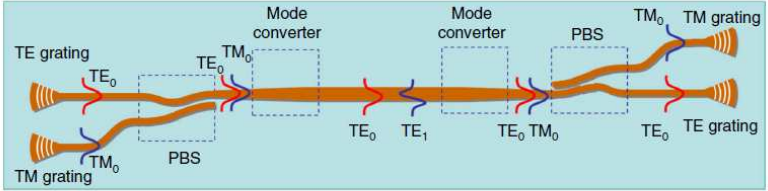
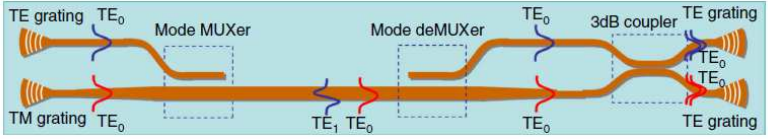
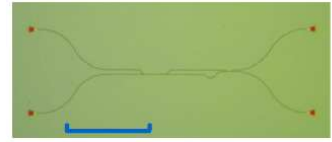
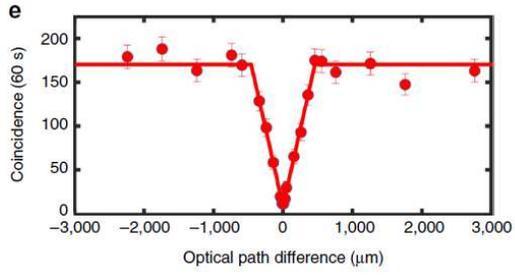
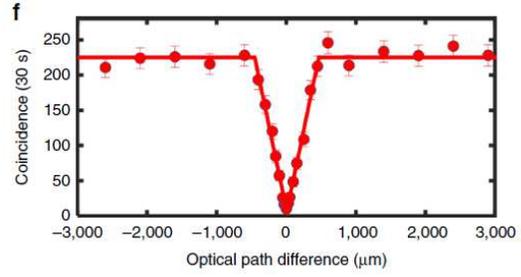
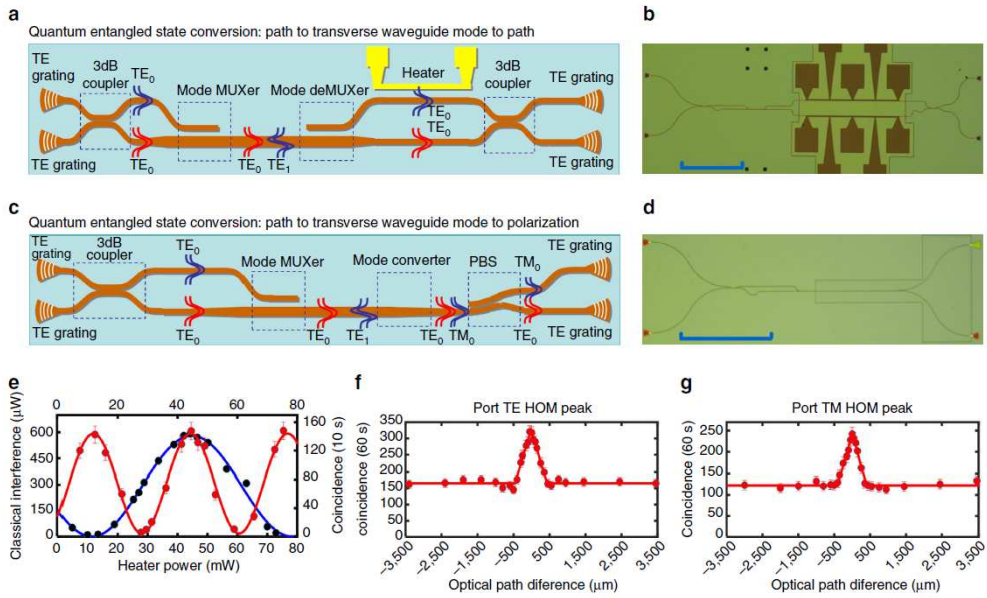
\end{document}